\documentclass[showpacs,preprintnumbers,amsmath,amssymb]{revtex4}

\usepackage{graphicx}
\usepackage{dcolumn}
\usepackage{bm}

\begin{document}

\title{The covariant and on-shell statistics in $\kappa$-deformed spacetime}

\author{Rong-Xin Miao}
\email{mrx11@mail.ustc.edu.cn} \affiliation{University of Science
and Technology of China\\Hefei, Anhui 230026, P.R.China }
\begin{abstract}
   It has been a long-standing issue to construct the statistics of identical particles in
 $\kappa$-deformed spacetime. In this letter, we investigate different ideas on this problem.
 Following the ideas of Young and Zegers, we obtain the covariant and on shell kappa two-particle
 state in $1+1$ $D$ in a simpler way. Finally, a procedure to get such state in higher dimension
 is proposed.
\end{abstract}
\maketitle

\section{INTRODUCTION}
   It is possible that geometry is noncommutative at planck scale because of the quantum
 gravitational effects. Noncommutative models, such as canonical and Lie algebra type
 noncommutativity, have been studied extensively. In particular, the $\kappa$-Minkowski
 spacetime [1] drew great interest due to its connection with double special relativity [2,3].

   Recently, field theory in $\kappa$-Minkowski spacetime was extensively studied in [4-11]
 with some encouraging progress. [6,12] succeeded to obtain Noether charges associated
 with $\kappa$-Poincare symmetries of classical fields, and [4,7] generalized the results to
 the quantum level and got the finite vacuum energy. Besides, [13,14,15] suggested
 that the $\kappa$-Poincare symmetry might emerge in the low-energy limit of $1+2$
 dimensional quantum gravity.
   However, the $\kappa$-field theory is difficult to be quantized because of its singular
 statistics. Although many investigations have been made on this problem [4,5,16-21],
 it is yet not able to construct a $\kappa$-multi-particle state that is both covariant
 and on shell. As emphasized in [17] , to keep the notion
 of identical particles same in all frames, the covariant condition is necessary.
 It seems ridiculous that the bosons in one frame become fermions in another frame. The other
 condition ``on-shell''is also necessary, otherwise as we will show in Sec.III, there will be infinite
 covariant solutions.

   In this paper we prove that the twisted statistics is both covariant
 and on shell in the noncommutative spacetime constructed from twisted
 Poincare algebra, but it is often off shell in a more general noncommutative
 spacetime. We get the covariant and on-shell $\kappa$-statistics in $1+1$ $D$
 in a different way from [18]. Our method is simpler and applicable to all the
 $\kappa$-Poincare basis.

  The paper is organized as follows. In Sec.II, after
 a brief review of the $\kappa$-Poincare Hopf algebra and twisted Poincare Hopf
 algebra, we build a frame to solve the statistics issue in $\kappa$-deformed spacetime.
 In Sec.III we compare and study previous ideas on this problem. In Sec.IV we investigate
 the covariant and on-shell statistics in $\kappa$-deformed spacetime in  $1+1$ $D$. Finally we
 conclude the paper with some discussions.
\section{PRELIMINARIES}
\subsection{$\kappa$-Poincare algebra}
  Symmetries in $\kappa$-Minkowski spacetime are described by
 $\kappa$-Poincare Hopf algebra $P_\kappa$, which is a deformation of usual
 Poincare Hopf algebra P. For example, the $\kappa$-statistics must be invariant
 under the action of $P_\kappa$. In this section, we first review the
 $\kappa$-Poincare Hopf algebra in the bicrossproduct basis [22], then
 show the difficulties to construct the corresponding $\kappa$-statistics.

  $\kappa$-Poincare Hopf algebra is as
 follows,

 (a) algebra sector
 \begin{equation}
 [N_i,P_0]=iP_i\nonumber
 \end{equation}
 \begin{eqnarray}
 [M_i,P_j]=i\epsilon_{ijk}P_k,\mbox{ }\mbox{ }[M_i,P_0]=0
 \end{eqnarray}
 \begin{equation}
[M_i,M_j]=i\epsilon_{ijk}M_k \nonumber\\
 \end{equation}
 \begin{eqnarray}
  [M_i,N_j]=i\epsilon_{ijk}N_k,\mbox{ }[N_i,N_j]=-i\epsilon_{ijk}M_k
 \end{eqnarray}
 \begin{equation}[N_i,P_j]=i\delta_{ij}[\frac{k}{2}(1-e^{-\frac{2P_0}{k}})+\frac{1}{2k}\overrightarrow{P}^2]-\frac{i}{k}P_iP_j
 \end{equation}

 (b) coalgebra
 $$\triangle P_0=P_0\otimes 1+1\otimes P_0$$
 \begin{eqnarray}
 \triangle M_i=1\otimes M_i+M_i\otimes 1\end{eqnarray}
 \begin{eqnarray}\triangle P_i=P_i\otimes 1+e^{-\frac{P_0}{k}}\otimes P_i
 \end{eqnarray}
 \begin{equation}\triangle N_i=N_i\otimes
 1+e^{-\frac{P_0}{k}}\otimes
 N_i+\frac{1}{k}\epsilon_{ijk}P_i\otimes M_k
 \end{equation}

 (c) antipodes
 \begin{equation}S(P_0)=-P_0,\mbox{ } \mbox{ } S(P_i)=-P_ie^{\frac{P_0}{k}}
 \end{equation}
 \begin{eqnarray}S(M_i)=-M_i, \mbox{
 }S(N_i)=-N_ie^{\frac{P_0}{k}}+\frac{1}{k}\epsilon_{ijk}P_je^{\frac{P_0}{k}}M_k\nonumber\\
 \end{eqnarray}

 (d) co-units
 $$\epsilon(P_u)=\epsilon(M_j)=\epsilon(N_i)=0$$
where the Casimir operator $C_k$ is
 \begin{equation}C_k=(2k\sinh{\frac{P_0}{2k}})^2-\overrightarrow{P}^2e^{\frac{P_0}{k}}
 \end{equation}

  It is worth noting that the coproduct $\triangle$ is deformed,
 which brings many new features in $\kappa$-Minkowski spacetime. For example,
 the addition law becomes nonlinear and non-abelian: $$p\oplus
 k=(p_0+k_0,\mbox{ }p_i+k_ie^{-\frac{p_0}{k}})$$
 \begin{equation}k\oplus p=(k_0+p_0,\mbox{ }k_i+p_ie^{-\frac{k_0}{k}})
 \end{equation}
  The oddness is due to the asymmetrical coproduct (5)
 \begin{eqnarray}&&P_i \triangleright (e^{ip\cdot x}\otimes e^{ik\cdot
 x})\nonumber\\
 &=&\triangle(P_i)(e^{ipx}\otimes e^{ikx})\nonumber\\
 &=&(P_i\triangleright e^{ipx})\otimes e^{ikx}+(e^{-\frac{P_0}{k}}\triangleright e^{ipx})\otimes(P_i\triangleright
 e^{ikx})\nonumber\\
 &=&(p_i+k_ie^{-\frac{p_0}{k}})(e^{ipx}\otimes e^{ikx})
 \end{eqnarray}
  While the asymmetrical coproduct is due to the noncommutative
 spacetime. Assume the common inner product$<x_u,P_v>=-i\eta_{uv}$, one can obtain
 that,
 \begin{eqnarray}&&<[x_i,x_0],P_k>\nonumber\\
 &=&<x_i\otimes
 x_0,\triangle(P_k)>-<x_0\otimes x_i,\triangle(P_k)>\nonumber\\
 &=&<x_i,P_k><x_0,e^{-\frac{P_0}{k}}>-<x_0,1><x_i,P_k>\nonumber\\
 &=&<i\frac{1}{k}x_i,P_k>\nonumber\\
 &&\Rightarrow [x_i,x_0]=i\frac{x_i}{k}
 \end{eqnarray}
  As we expect, it is just the $\kappa$-Minkowski spacetime.

  For the non-abelian addition law between momentum (10) , the usual statistics
 \begin{equation}\mid p>\otimes\mid q>\pm \mid q>\otimes \mid p>
 \end{equation}
 fail in $\kappa$-Minkowski spacetime. Because it is no longer the
 eigenstates of momentum. This is one of the difficulties which one
 encounters in obtaining the statistics in $\kappa$-deformed spacetime.

 \subsection{Twisted Poincare algebra}
 From the Poincare Hopf algebra, one can construct a
 new Hopf algebra if there exists a twist element
 $F\in U(g)\otimes U(g)$, which satisfies the countital 2-cocyle condition.
 \begin{equation}(F\otimes 1)(\triangle\otimes id)F=(1\otimes
 F)(id\otimes\triangle)F
 \end{equation}
 \begin{equation}(\epsilon\otimes id)F=1=(id\otimes \epsilon)F
 \end{equation}
  It is known that F does not modify the counit and the algebra part,
 but changes the coproducts and the antipodes.
 \begin{equation}\triangle_F(g)=F\triangle_0(g)F^{-1}
 \end{equation}
 \begin{equation}S_F(g)=US_0(g)U^{-1}
 \end{equation}
 \begin{equation}U=\sum F_{(1)}S(F_{2})
 \end{equation}
 Let A be the algebra on which $U(g)$ acts, then one can define the
 star product
 \begin{equation}f*g=m_0(F^{-1}\triangleright(f\otimes g))
 \end{equation}
for $f,g\in A$.

 It is interesting that the canonical noncommutativity can be
 constructed from twisted Poincare Hopf algebra with the twist element
 \begin{equation}F_\theta=exp(\frac{i}{2}\theta^{\alpha\beta}P_{\alpha}\otimes P_{\beta})
 \end{equation}
 From(16)-(20), one can get:
 $$[x_u,x_v]=i\theta_{uv}$$
 $$\triangle_\theta(P_u)=\triangle_0(P_u)$$
 \begin{eqnarray}&&\mbox{ }\mbox{ }\mbox{ }\mbox{ }\mbox{ }\mbox{ }\mbox{ }\mbox{ }\triangle_\theta(M_{uv})=\triangle_0(M_{uv})-\nonumber\\
 &&\frac{i}{2}\theta^{\alpha\beta}((\eta_{\alpha u}P_v-\eta_{\alpha v}P_u)\otimes
 P_\beta+(\alpha \leftrightarrow \beta))
 \end{eqnarray}

   It is necessary to point out that one is not able to derive the $\kappa$-Poincare algebra from
 twisted Poincare Hopf algebra. Instead, one need a larger algebra,
 for example, the Hopf algebra  $ifl(n,R)$ [23].
 And the corresponding twist element is
 \begin{equation}F_k=exp[\frac{i}{2k}(P_0\otimes D-D\otimes P_0)]
 \end{equation}
 where $D=x_iP_i$ stands for dilation.
\subsection{The frame to solve $\kappa$-Statistics}
 To get the covariant and on-shell statistics, we just need to
 construct a proper flip operator $\tau$.
 \begin{equation}\tau:\mid p>\otimes\mid q>\rightarrow\mid \widetilde{q}>\otimes
 \mid\widetilde{p}>
 \end{equation}
Then the two-particle state becomes
 \begin{equation}\mid p,q>=\frac{1}{\sqrt{2}}(1\pm\tau)\mid
 p>\otimes\mid q>
 \end{equation}
 with +($-$) for bosons (fermions).

 Take the commutative case as an
 example, the proper flip operator $\tau_0$ must satisfy:
 \begin{equation}[\tau_0,\triangle(g)]=0
 \end{equation}
 \begin{equation}[\tau_0,C_0\otimes 1]=[\tau_0,1\otimes C_0]=0
 \end{equation}
 \begin{equation}\tau_0^2=1
 \end{equation}
for $\forall g\in P$. Condition (25) means that the two-particle
state (24) is the momentum eigenstate and the statistics is
covariant. $C_0$ is the Casimir operator, so condition (26) means
that all the identical particles are on shell. While the requirement
(27) leads to
 \begin{equation}\tau \mid p,q>=\pm\mid p,q>
 \end{equation}

  Obviously, the flip operator $\tau_0$ that satisfies the above
conditions is just the common exchange:
 \begin{equation}\tau_0\mid p>\otimes\mid q>=\mid q>\otimes\mid p>
 \end{equation}
  However, as is mentioned in Sec.II.A, $\tau_0$ is no longer the
solution of
 $\kappa$-statistics since (13) is not the momentum eigenstate. So we have to
 search for a new $\tau_\kappa$ over again with the similar conditions,
 \begin{equation}[\tau_\kappa,\triangle_\kappa(g)]=0
 \end{equation}
 \begin{equation}[\tau_\kappa,C_\kappa\otimes 1]=[\tau_\kappa,1\otimes C_\kappa]=0
 \end{equation}
 \begin{equation}\tau_\kappa^2=1
 \end{equation}

  Before we begin to seek for $\tau_\kappa$, let us first
 simplify these conditions as much as possible.
 From the equations (1)(2)
 and identity $\triangle(ab)=\triangle(a)\triangle(b)$ , we can derive
 \begin{equation}[\triangle(N_i),\triangle(N_j)]=-i\epsilon_{ijk}\triangle(M_k)
 \end{equation}
 \begin{equation}[\triangle(N_i),\triangle(P_0)]=i\triangle(P_i)
 \end{equation}
 So instead of the conditions (30) for all the generators of
 $\kappa$-Poincare algebra, we only need
 \begin{equation}[\tau_k,\triangle_k(N_i)]=0
 \end{equation}
 \begin{equation}[\tau_k,\triangle_k(P_0)]=0
 \end{equation}

 Once we have found the solution of $\tau_k$, we can construct
 the $\kappa$-Fock space. However, in this paper we only focus on the two-particle Hilbert space
 :
 \begin{equation}\mid p,q>=\frac{1}{\sqrt{2}}(1\pm\tau_k)\mid
 p>\otimes\mid q>
\end{equation}
As to the whole Fock space, please refer to [18,19,21] for more
details.

 From (37), one can get the oscillator algebras in
 $\kappa$-Minkowski spacetime.
 \begin{eqnarray}[a_p^+,\mbox{ }a_q^+]_\kappa&=&a_{p}^{+}a_{q}^{+}-\tau_\kappa(a_{p}^{+}a_{q}^{+})\nonumber\\
 &=&a_{p}^{+}a_{q}^{+}-a_{\widetilde{q}}^{+}a_{\widetilde{p}}^{+}=0
 \end{eqnarray}
 The definition of $\widetilde{q}$ and $\widetilde{p}$ is as follows:
 \begin{equation}\tau_k\mid p>\otimes\mid q>=\mid
 \widetilde{q}>\otimes\mid \widetilde{p}>
 \end{equation}
 As to the other algebra relations, we only need to replace
$a_p$ with $a_{S(p)}^{+}$. Then we are able to calculate the Feynman
propagator, Pauli-Jordan commutator function and so on.

\section{VARIOUS IDEAS ABOUT $\kappa$-STATISTICS}
\subsection{Twisted Statistics in $\kappa$-Minkowski spacetime}
   Statistics in canonical noncommutative spacetime have been solved
 ideally by the twisted method. And this method is generalized to $\kappa$-Minkowski
 spacetime in [16]. We first review the main ideas and results in [16].
 Then we prove that the twisted statistics in [16] is off shell and
 it is a general conclusion for the noncommutativity that can not be
 derived from twisted Poincare algebra.

   Consider a system with symmetry group G in commutative spacetime.
 Let $\Lambda$ be an element of G which acts with some
 representation D. The action of G on the two-particle Hilbert space
 is described by the coproduct $\triangle_0$:
 $$\triangle_0:\lambda\rightarrow\lambda\otimes\lambda$$
 \begin{equation}f\otimes g\rightarrow(D\otimes
 D)\triangle_0(\lambda)f\otimes g
 \end{equation}

  To be consistent with the usual multiplication map
 $m_0$, the coproduct must satisfy the condition
 \begin{equation}m_0((D\otimes D)\triangle_0(\lambda)f\otimes g)=D(\lambda)m_0(f\otimes g)
 \end{equation}
 Because the multiplication map $m_0$ is abelian, the coproduct
 $\triangle_0$ must be symmetrical. Thus to satisfy the covariant
 condition
 \begin{equation}[\triangle_0,\tau_0]=0
 \end{equation}
 one only need to define  $\tau_0$ as the usual exchange:
 \begin{equation}\tau_0(f\otimes g)=g\otimes f
 \end{equation}
 And one can easily check that the conditions (25)-(27)are satisfied
 automatically.

 In the noncommutative spacetime, the non-abelian star product is
 defined with the twist element F as
 \begin{equation}f*g=m_0(Ff\otimes g)=m_k(f\otimes g),m_k=m_0F
 \end{equation}
 For example, for any $\varphi$ ordering [24] in $\kappa$-Minkowski
 spacetime, the corresponding twist element $F_\varphi$ is
 \begin{eqnarray}F_\varphi=exp{(N_x[ln\phi(A_x+A_y)-ln\phi(A_x)]+(x\leftrightarrow y))}
 \end{eqnarray}
 where $N_x=x_i\partial /\partial x_i$ and similarly for
 $N_y$, $A_x=ia\partial_n^x$ and similarly for $A_y$ [16].

 For the non-abelian multiplication $m_\varphi=m_0F_\varphi$, one has
 to redefine the coproduct $\triangle_\varphi$ as
 \begin{equation}\triangle_\varphi=F_\varphi^{-1}\triangle_0F_\varphi
 \end{equation}
 in order to satisfy the equation
 \begin{equation}m_\varphi [(D\otimes D)\triangle_\varphi(\lambda)f\otimes
 g]=D(\lambda)m_\varphi(f\otimes g)
 \end{equation}
 However, it turns out that the flip operator $\tau_0$ does not
 commute with the twisted coproduct. One must find a new flip operator
 $\tau_\varphi$ to be consistent with the covariant condition.
 \begin{equation}[\triangle_\varphi,\tau_\varphi]=0
 \end{equation}
 The simplest solution of $\tau_\varphi$ is
 \begin{equation}\tau_\varphi=F_\varphi^{-1}\tau_0F_\varphi
 \end{equation}
 From(45)-(49), the authors of [16] found that for the particular
 class of $\varphi$ realizations, the twisted flip operator
 is independent of the choice of ordering
 \begin{equation}\tau_\varphi=exp{[i(x_iP_i\otimes A-A\otimes
 x_iP_i)]}\tau_0
 \end{equation}

 Now let us do some discussions about the main result (50) in [16].
 First of all, one may note that (49) is a sufficient but not
 a necessary condition of (48). In fact, for any non-degenerative
 operator F which satisfies the following condition
 \begin{equation}[\triangle_\varphi,F]=F
 \end{equation}
 \begin{equation}\tau'_\varphi=F^{-1}\tau_\varphi F
 \end{equation}
 $\tau'_\varphi$ is also one of the solutions of (48). One can not
 simply choose (49).

 Secondly, we find that the flip operator (50) is off shell.

 Proof:
 \begin{equation}N=x_i\partial_i,\mbox{ } [\partial_i,x_j]=\delta_j^i,\mbox{ } [A,N]=0
 \end{equation}
 The symbol : : is the normal ordering with all $x_i$ coming to the
 left of all $\partial_i$.
 According to the appendix of [24], we have
 \begin{equation}:e^{N(e^{A}-1)}:=e^{NA}
 \end{equation}
 so
 \begin{eqnarray}\tau_\varphi|p>\otimes|k>&=&e^{iN\otimes
 A}\tau_0e^{-iN\otimes A}|p>\otimes|k>\nonumber\\
 &=&e^{iN\otimes A}\tau_0:e^{N(e^{-i\otimes
 A}-1)}:|p>\otimes|k>\nonumber\\
 &=&e^{iN\otimes A}\tau_0|p_0,p_ie^{-ak_0}>\otimes|k>\nonumber\\
 &=&e^{iN\otimes A}|k>\otimes|p_0,p_ie^{-ak_0}>\nonumber\\
 &=&|k_0,k_ie^{ap_0}>\otimes|p_0,k_ie^{-ak_0}>
 \end{eqnarray}

 In the above calculations we use $a=1/\kappa$ and the equation
 \begin{equation}e^{-iA\otimes N}\tau_0=\tau_0e^{-iN\otimes A}
 \end{equation}
 Because $(k_0, k_i)$ and $(p_0, p_i)$ are both on
 shell, and k and p are irrelevant , so
 $(k_0, k_ie^{ap_0})$ and $(p_0, p_ie^{-ak_0})$ must be both off shell.
 Now the proof is completed.

 Last but not least, let us seek for the deep reason why the twisted statistics succeed
 in canonical noncommutative spacetime but fail in $\kappa$-Minkowski spacetime. As is
 mentioned in section II.C, the key lies in the fact that the canonical noncommutativity
 can be constructed from twisted Poincare Hopf algebra but
 $\kappa$-Minkowski spacetime can not.

 For the canonical case, $C_\theta(P)=C_0(P)=P_\mu P^\mu$ $$F_\theta=exp{\frac{i}{2}\theta^{\alpha\beta}P_\alpha\otimes P_\beta}$$
 \begin{equation}[C_\theta(P)\otimes 1,F_\theta]=[1\otimes
 C_\theta(P),F_\theta]=0
 \end{equation}
 so it is easy to derive the on-shell condition
 \begin{equation}[C_\theta(P)\otimes 1,\tau_\theta]=[1\otimes
 C_\theta(P),\tau_\theta]=0
 \end{equation}

 It implies that for all kinds of noncommutativities which can be
 constructed from twisted Poincare algebra, the twisted
 statistics is both covariant and on shell. That is because
 $$C_F(P)=C_0(P)=P_\mu P_\mu$$
 \begin{equation}[C_F(P)\otimes 1,F]=[1\otimes C_F(P),F]=0
 \end{equation}

 As to the case in $\kappa$-Minkowski spacetime, the twisted element $F_k$ of
 a bigger algebra must contain the dilation operator D . Notice that
 \begin{equation}\forall f(P)\neq 0,[f(P),D]=\frac{\partial f(P)}{\partial
 P_\mu}P_\mu\neq 0
 \end{equation}
  so the twisted statistics must be off shell .$$[C_k(P)\otimes 1,F_k]\neq 0$$
 \begin{equation}[C_k(P)\otimes 1,\tau_k]\neq 0
 \end{equation}

 It implies that for the most general kinds of noncommutativities which
 can not be derived from the twisted Poincare algebra, the twisted statistics will fail if the twisted
 element F contains generator that dose not commutate with $C(P)$.

\subsection{Rainbow Statistics}
 In a recent paper [21], the authors named the $\kappa$-statistics
 an romantic name---``Rainbow statistics.'' In [21], they treat the massive
 and massless case differently. For massive fields they got the covariant but
 off-shell $\kappa$-statistics by the twisted method, while for massless
fields [4,21] they
 got the on-shell statistics using the $\tau_k$ ,
 $$\tau_\kappa|p>\otimes |q>=|\widetilde{q}\otimes|\widetilde{p}>$$
 $$\widetilde{q}=(\widetilde{q_0},q_ie^{-\frac{p_0}{\kappa}}),\widetilde{p}=(\widetilde{p_0},p_ie^{\frac{\widetilde{q_0}}{\kappa}})$$
 \begin{equation}\widetilde{q_0}=-\kappa ln(1-\frac{|\overrightarrow{q}|}{\kappa}e^{-\frac{p_0}{\kappa}}),\widetilde{p_0}=-\kappa ln(1-\frac{|\overrightarrow{p}|}{\kappa}e^{\frac{\widetilde{q_0}}{\kappa}})
 \end{equation}
 where $\widetilde{q}$ and $\widetilde{p}$ are both on shell, and the
 Casimir operator is (9). We can easily check that
 \begin{equation}\tau_\kappa^2=1, [\tau_\kappa,\triangle(P_\mu)]=0
 \end{equation}

 But we find that the $\kappa$-statistics constructed from $\tau_k$ (62) is not covariant.
 Now we give a simple proof. As discussed in Section II.B, we only need
 to check if $\triangle(N_i)$ commute with $\tau_\kappa$. For
 simplicity , we consider an infinitesimal Lorentz transformation in
 $1+1$ $D$.
 $$1+i\alpha N_1,\mbox{ }\mbox{ }\alpha\rightarrow 0$$
 with algebra sector and coproduct as,
 $$[P_0,N_1]=-iP_1,\mbox{ }[P_1,N_1]=-i\frac{\kappa}{2}(1-e^{-\frac{2P_0}{\kappa}})-\frac{1}{2\kappa}P_1P_1$$
 \begin{equation}\triangle N_1=N_1\otimes
 1+e^{-\frac{P_0}{\kappa}}\otimes N_1
 \end{equation}

 Following the method in [17] , we get
 \begin{eqnarray}&&(1\otimes 1+\alpha\triangle
 N_1)|p_0,p_1>\otimes |q_0,q_1>\nonumber\\
 &=&|p_0-\alpha
 p_1,p_1-\alpha[\frac{\kappa}{2}(1-e^{-\frac{2p_0}{\kappa}}-\frac{1}{2\kappa}p_1p_1)]>\nonumber\\
 &\otimes &|q_0-\alpha e^{-\frac{p_0}{k}}q_1,q_1-\alpha
 e^{-\frac{p_0}{k}}[\frac{k}{2}(1-e^{-\frac{2q_0}{k}})-\frac{1}{2k}q_1q_1]>\nonumber\\
 \end{eqnarray}
Define
\begin{eqnarray}&&|Q_0,Q_1>\otimes |P_0,P_1>\nonumber\\
&=&(1\otimes 1+\alpha\triangle N_1)\tau_k|p_0,p_1>\otimes
|q_0,q_1>\nonumber\\
&&|\overline{Q_0},\overline{Q_1}>\otimes|\overline{P_0},\overline{P_1}>\nonumber\\
&=&\tau_k(1\otimes 1+\alpha\triangle
 N_1)|p_0,p_1>\otimes|q_0,q_1>
 \end{eqnarray}
 then after complicated calculations, we get
 \begin{eqnarray}&&Q_1-\overline{Q_1}\nonumber\\
 &=&q_1e^{-\frac{p_0}{k}}-\alpha|q_1|e^{-\frac{p_0}{k}}+\alpha\frac{q_1^2}{k}e^{-\frac{2p_0}{k}}-q_1e^{-\frac{p_0}{k}}+O(\alpha^2)\nonumber\\
 &-&\alpha\frac{q_1|p_1|}{k}e^{-\frac{p_0}{k}}+\frac{\alpha
 k}{2}e^{-\frac{2p_0}{k}}(1-e^{-\frac{2q_0}{k}})-\frac{\alpha
 q_1^2}{2k}e^{-\frac{2p_0}{k}}\nonumber\\
 &=&-\alpha(\frac{|q_1||p_1|}{k}+\frac{q_1p_1}{k}+\frac{p_1^2|q_1|}{k^2}+\frac{p_1|p_1|q_1}{k^2})+O(\alpha^2)\nonumber\\
 \end{eqnarray}

 Obviously, when $q_1p_1>0 , Q_1-\overline{Q_1}\neq 0$. Only in the case $q_1p_1<0$ , $Q_1-\overline{Q_1}=0$. So the massless
 Rainbow Statistics in [4,21] is not covariant.
 \subsection{Other $\kappa$-Statistics}
 There are also some other attempts to construct $\kappa$-statistics in
 $\kappa$-deformed spacetime, now we only give a brief introduction.
  In [5,20], the authors introduced a new $\kappa$-star product and
 the corresponding oscillator algebra in order to get the full
 Fock space of $\kappa$-quantum field theory. It is
 interesting that the classical four-momentum conservation law is satisfied
 in their scenario. However, they modified the on-shell conditions,
 and did not consider the covariant conditions .
  In [17], the authors succeeded to get the covariant and on-shell
 $\kappa$-statistics to the third order in $1/\kappa$. Then in the
 following work [18], they obtained the exact solution in 1+1 $D$.
  For more details of these papers, one can refer to the references [4,21,5,10,17,18].

\section{THE COVARIANT AND ON-SHELL $\kappa$-STATISTICS IN 1+1 D}
 It is argued in [18] that the covariant and on- shell
 $\kappa$-statistics in $1+1$ $D$ has been found. It is also
 argued that for the case of two-particle state, their realization is
 unique. However, the elliptic functions are contained in their result.
 In this section, we solve the same problem in a quite different way,
 our method is simpler and our result contains only elementary functions.

 As is emphasized in Section II.C, our goal is to find a proper $\tau_k$
 which satisfies the equations:
 $$\tau_k|p_0,p_1>\otimes|q_0,q_1>\rightarrow|\widetilde{q_0},\widetilde{q_1}>\otimes|\widetilde{p_0},\widetilde{p_1}>$$
 $$[\tau_k,\triangle(N_1)]=0,\mbox{ }[\tau_k,\triangle(p_0)]=0$$
 $$[C_k\otimes 1,\tau_k]=[1\otimes C_k,\tau_k]=0$$
 \begin{equation}\tau_k^2=1
 \end{equation}
 As we will show later, the condition $\tau_k^2=1$ is satisfied automatically.
 Now there are four unknown quantities with four independent equations. So the
 solution of $\tau_k$ in $1+1$ $D$ is unique.

 In view of the uniqueness, then we can replace the condition
 $[\tau_k,\triangle(N_1)]=0$ with
 $[\tau_k,\triangle(p_1)]=0$. Now the equations (68) become
 $$p_0+q_0=\widetilde{q_0}+\widetilde{p_0}$$
 $$p_1+q_1e^{-\frac{p_0}{k}}=\widetilde{q_1}+\widetilde{p_1}e^{-\frac{\widetilde{q_0}}{k}}$$
 \begin{equation}C_k(q_0,q_1)=C_k(p_0,p_1)=C_k(\widetilde{q_0},\widetilde{q_1})=C_k(\widetilde{p_0}\widetilde{p_1})=m^2
 \end{equation}
 where $C_k$ is the Casimir operoter (9). Simplify the above
 equations, one can easily get the relations between
 $\widetilde{p_1}$ and $\widetilde{q_1}$:
 \begin{eqnarray}&&p_1+q_1(M-\sqrt{M^2-1+\frac{p_1^2}{\kappa^2}})\nonumber\\
 &=&\widetilde{q_1}+\widetilde{p_1}(M-\sqrt{M^2-1+\frac{\widetilde{q_1}^2}{\kappa^2}})
 \end{eqnarray}
 \begin{eqnarray}&&(M-\sqrt{M^2-1+\frac{p_1^2}{\kappa^2}})(M-\sqrt{M^2-1+\frac{q_1^2}{\kappa}})\nonumber\\
 &=&(M-\sqrt{M^2-1+\frac{\widetilde{q_1}^2}{\kappa^2}})(M-\sqrt{M^2-1+\frac{\widetilde{p_1}^2}{\kappa^2}})\nonumber\\
 \end{eqnarray}
 where $M=1+m^2/2\kappa^2$. Obviously, one solution of
 (70)(71) is
 \begin{equation}\widetilde{q_1}=p_1,\mbox{  } \mbox{  }\mbox{  }\mbox{  }\mbox{  }\mbox{  }\mbox{  }\widetilde{p_1}=q_1
 \end{equation}
 However, it is not the solution we want. Eliminating $\widetilde{p_1}$
 in (70)(71),  we find that the equation of $\widetilde{q_1}$ is
 just an quadratic equation. Having known one of the solution, we can
 easily get the other one:
 \begin{eqnarray}\widetilde{q_1}&=&(b-p_1)-\frac{b(a^2-1)(M^2-1)}{M^2(a-1)^2-\frac{b^2}{\kappa^2}}\nonumber\\
 &=&q_1e^{-\frac{p_0}{k}}-\frac{b(a^2-1)(M^2-1)}{M^2(a-1)^2-\frac{b^2}{\kappa^2}}\nonumber\\
 \widetilde{p_1}&=&(b-\widetilde{q_1})e^{\frac{\widetilde{q_0}}{k}}
 \end{eqnarray}
 with
 $$b=p_1+q_1(M-\sqrt{M^2-1+\frac{p_1^2}{\kappa^2}})$$
$$a=(M-\sqrt{M^2-1+\frac{p_1^2}{\kappa^2}})(M-\sqrt{M^2-1+\frac{q_1^2}{\kappa^2}})$$
 Now let us do some discussions about the above results.

 First of all, it is worth noting that the flip operator $\tau_k$ is just the
 exchange between the two solutions (72) and (73) of the equations
 (70)(71). So the condition $\tau_k^2=1$ comes to be true automatically.

 Secondly, the $\tau_k$ we get in (73) must be covariant. In general,
 the equations (69) are only the necessary conditions of
 (68). But in our case, there is only one solution which satisfy
 (68). So the conditions (68) and (69) must be equivalent to each
 other.

 Finally, notice that the first part of $\widetilde{q_1}$ in (73) is just the result got in
 [4,21] for massless fields. In the massless limit $(M\rightarrow
 1)$, our result becomes
 \begin{equation}\widetilde{q_1}=q_1e^{-\frac{p_0}{\kappa}}=q_1(1-\frac{|p_1|}{\kappa}),\mbox{ } p_1q_1<0
\end{equation}
\begin{equation}\widetilde{q_1}=q_1(1-\frac{|p_1|}{\kappa})-\frac{c}{d},\mbox{ }p_1q_1>0\nonumber\\
\end{equation}
where
\begin{eqnarray}
 c=(p_1+q_1-\frac{q_1|p_1|}{\kappa})(2-\frac{|p_1|}{\kappa}-\frac{|q_1|}{\kappa}+\frac{|p_1q_1|}{\kappa^2})\nonumber\\
 d=(2-\frac{2|p_1|}{\kappa}-\frac{|q_1|}{\kappa}-\frac{\kappa}{|q_1|}+\frac{|p_1|}{q_1}+\frac{|q_1p_1|}{\kappa^2})
\end{eqnarray}

 For the case (74), we have proved in Section III.B, it is covariant.
 While for the other case (75), following the same program, after
 complex calculation we find exactly that it is indeed covariant
 too.

\section{SUMMARY AND OUTLOOK}
  In this paper, we have investigated the statistics problem in
 $\kappa$-Minkowski spacetime. Focusing on the two-particle state,
 we have obtained the covariant and on-shell $\kappa$ two-particle states in $1+1$
 $D$ using a simpler method than [18]. Our result contains only elementary functions and applicable to all the
 $\kappa$-Poincare basis. Now, we suggest a scheme
 to construct the covariant and on-shell $\kappa$-two particle states in
 higher dimension. Notice that one can rotate an arbitrary
 $\kappa$-two particle state in (1+n) D into (1+1) D . One idea
 quickly comes to mind. One may first rotate the $\kappa$ two-particle state in $(1+n)$ $D$
 into $(1+1)$ $D$, then after the action of $\tau_k^{(1)}$ which have been obtained,
 one rotate it back into $(1+n)$ $D$. Whether this method is
 workable and how to carry out the scheme explicitly are open
 problems.

   We end the paper with some problems. Now there are too many
 different kinds of noncommutativity, but there is only one
 world. So we must search for some principle to restrict the
 possible form of noncommutativity.
 Just like in the field theory, the gauge invariance can largely
 restrain the form of interaction. Of course, which kind of noncommutativity
 is right depends on the experiments. But if noncommutativity is
 a correct and also beautiful theory, it must be able to give some
 predictions in theory. Besides, there may exist duality between
 different kinds of noncommutativity. Whether it does exist and how
 to seek such duality is also an open problem.
 Hopefully that some insights can emerge in the future.

\section*{Acknowledgements}

  We are grateful to Miao Li and Yan-Gang Miao for kind help and suggestions .
 We thank Tianming Bird for useful discussions.

\end{document}